\documentclass[aps,pra,preprint,superscriptaddress]{revtex4-1}

\usepackage{graphicx}
\usepackage{amsmath}

\begin{document}

\title{Sub-microsecond entangling gate between trapped ions via Rydberg interaction}

\author{Chi Zhang}
\email[]{chi.zhang@fysik.su.se}
\affiliation{Department of Physics, Stockholm University, 10691 Stockholm, Sweden}
\author{Fabian Pokorny}
\affiliation{Department of Physics, Stockholm University, 10691 Stockholm, Sweden}
\author{Weibin Li}
\affiliation{School of Physics and Astronomy, University of Nottingham, Nottingham, NG7 2RD, United Kingdom}
\affiliation{Centre for the Mathematics and Theoretical Physics of Quantum Non-equilibrium Systems, University of Nottingham, Nottingham, NG7 2RD, United Kingdom}
\author{Gerard Higgins}
\affiliation{Department of Physics, Stockholm University, 10691 Stockholm, Sweden}
\author{Andreas P\"oschl}
\affiliation{Department of Physics, Stockholm University, 10691 Stockholm, Sweden}
\author{Igor Lesanovsky}
\affiliation{School of Physics and Astronomy, University of Nottingham, Nottingham, NG7 2RD, United Kingdom}
\affiliation{Centre for the Mathematics and Theoretical Physics of Quantum Non-equilibrium Systems, University of Nottingham, Nottingham, NG7 2RD, United Kingdom}
\affiliation{Institut f\"ur Theoretische Physik, Universit\"at T\"ubingen, Auf der Morgenstelle 14, 72076 T\"ubingen, Germany}
\author{Markus Hennrich}
\email[]{markus.hennrich@fysik.su.se}
\affiliation{Department of Physics, Stockholm University, 10691 Stockholm, Sweden}

\begin{abstract}
Generating quantum entanglement in large systems on time scales much shorter than the coherence time is key to powerful quantum simulation and computation. Trapped ions are among the most accurately controlled and best isolated quantum systems \cite{wineland1998} with low-error entanglement gates operated via the vibrational motion of a few-ion crystal within tens of microseconds \cite{ballance2016}. To exceed the level of complexity tractable by classical computers the main challenge is to realise fast entanglement operations in large ion crystals \cite{zhang2017,hempel2018}. The strong dipole-dipole interactions in polar molecule \cite{anderegg2018} and Rydberg atom \cite{bernien2017,ravets2014} systems allow much faster entangling gates, yet stable state-independent confinement comparable with trapped ions needs to be demonstrated in these systems \cite{saffman2016}. Here, we combine the benefits of these approaches: we report a $700\,\mathrm{ns}$ two-ion entangling gate which utilises the strong dipolar interaction between trapped Rydberg ions and produce a Bell state with $78\%$ fidelity. The sources of gate error are identified and a total error below $0.2\%$ is predicted for experimentally-achievable parameters. Furthermore, we predict that residual coupling to motional modes contributes $\sim 10^{-4}$ gate error in a large ion crystal of 100 ions. This provides a new avenue to significantly speed up and scale up trapped ion quantum computers and simulators.
\end{abstract}

\maketitle

Trapped atomic ions are one of the most promising architectures for realizing a universal quantum computer \cite{wineland1998}. The fundamental single- and two-qubit quantum gates have been demonstrated with errors less than 0.1\% \cite{ballance2016}, sufficiently low for fault-tolerant quantum error-correction schemes \cite{fowler2012}. Nevertheless, a scalable quantum computer requires a large number of qubits and a large number of gate operations to be conducted within the coherence time. Most established gate schemes using a common motional mode are slow (typical gate times are between 40 and 100$\,\mu \mathrm{s}$) and difficult to scale up since the motional spectrum becomes more dense with increasing ion number. Many new schemes have been proposed \cite{garciaripoll2003, duan2004, garciaripoll2005, palmero2017}, with the fastest experimentally-achieved gate being 1.6$\,\mu \mathrm{s}$ (99.8\% fidelity) and 480\,ns (60\% fidelity) \cite{schafer2018}, realised by driving multiple motional modes simultaneously. Although the gate speed is not limited by the trap frequencies, the gate protocol requires the phase-space trajectories of all modes to close simultaneously at the end of the pulse sequence \cite{schafer2018}. In long ion strings with a large number of vibrational modes, it becomes increasingly challenging to find and implement laser pulse parameters that execute this gate with a low error. Thus, a slow-down of gate speed appears inevitable.

Two-qubit entangling gates in Rydberg atom systems are substantially faster, owing to strong dipole-dipole interactions. The gate fidelities in recent experiments using neutral atoms are fairly high \cite{levine2019,graham2019}. However, the atom traps need to be turned off during Rydberg excitation. This can cause unwanted coupling between qubits and atom motion as well as atom loss \cite{saffman2016,maller2015}. Employing blue-detuned optical tweezers at a magic wavelength one may achieve the trapping of Rydberg states \cite{piotrowicz2013}, though the predicted residual change in trapping frequency of $\sim 50\%$ \cite{zhang2011} will still result in entanglement between qubits and motional states. This challenge may set a limit on the achievable gate error.

Combining the benefits of trapped ion qubits and Rydberg interactions is a promising approach for scalable quantum computation \cite{mueller2008}. In previous works it was shown that ions in Rydberg state can be confined \cite{feldker2015, higgins2017_1} and coherence between Rydberg and low-lying states can be maintained \cite{higgins2017_2} in radio-frequency traps. However, strong interactions between Rydberg ions and their use for fast entangling gates had not been previously demonstrated.

\begin{figure}
	\includegraphics[width=\textwidth]{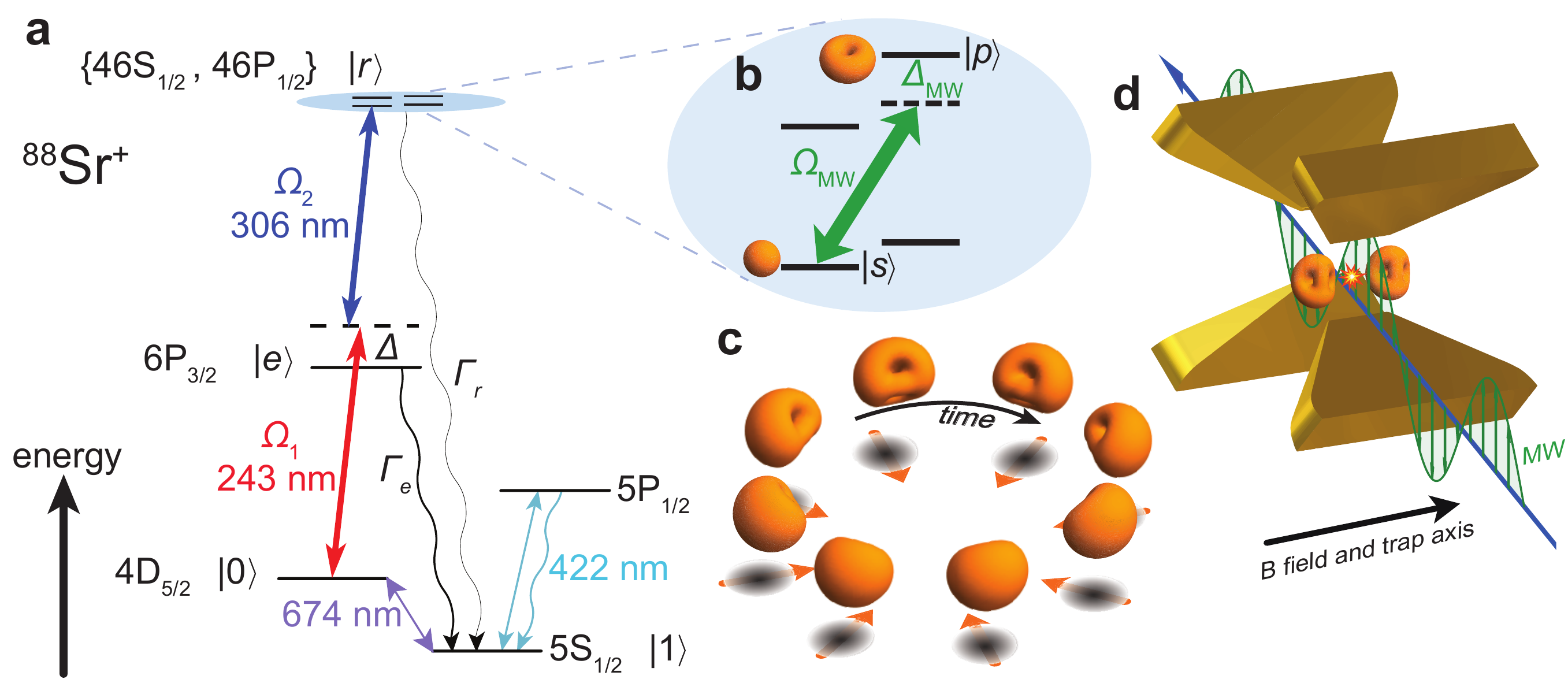}
	\caption{Level scheme of $^{88}$Sr$^+$ and the rotating dipole moment of a MW-dressed Rydberg state. (a) Ground state $5S_{1/2}, m_J=-1/2$ and metastable state $4D_{5/2}, m_J=-5/2$ are used to store a qubit $\{|1\rangle,|0\rangle\}$. The qubit transition is driven by a $674\,\mathrm{nm}$ laser field. Projective measurements in any basis are carried out by qubit rotations and fluorescence detection at $422\,\mathrm{nm}$. Rydberg excitation is driven by a two-photon laser field; $243\,\mathrm{nm}$ laser field couples $|0\rangle \leftrightarrow |e\rangle$ ($6P_{3/2}, m_J=-3/2$) and $306\,\mathrm{nm}$ laser field couples $|e\rangle \leftrightarrow |s\rangle$ ($46S_{1/2}, m_J=-1/2$). (b) A MW field at $122\,\mathrm{GHz}$ couples Rydberg state $|s\rangle$ with $|p\rangle$ ($46P_{1/2}, m_J=1/2$). (c) The Rydberg electron density for the MW-dressed state $\frac{1}{\sqrt{2}}(|s\rangle+|p\rangle)$ yields a permanent dipole which rotates in phase with the MW electric field. (d) Two ions are confined on the trap axis, their dipole moments rotate in-phase with the MW field about the magnetic field and the trap axis. The two MW-dressed Rydberg ions interact via the dipole-dipole interaction.}
	\label{Fig1}
\end{figure}

In our experiment $^{88}$Sr$^+$ ions are confined in a linear Paul trap. Two low-lying electronic states ($|0\rangle$ and $|1\rangle$) are used to store a qubit, and $|0\rangle$ is coupled to Rydberg state $|r\rangle$ via a two-photon laser field. The relevant level scheme is shown in Fig.~\ref{Fig1}(a), more details can be found in \cite{higgins2017_2}.

Two ions excited to Rydberg states interact through the dipole-dipole interaction
\begin{equation}
\hat{V}_{\mathrm{dd}} = \frac{1}{4\pi\epsilon_0} \left( \frac{\mathbf{\hat{\boldsymbol{\mu}}_1}\cdot\boldsymbol{\hat{\mu}_2} - 3(\boldsymbol{\hat{\mu}_1}\cdot\mathbf{n})(\boldsymbol{\hat{\mu}_2}\cdot\mathbf{n})}{|\mathbf{r}| ^3} \right)
\label{dd_interaction}
\end{equation}

where $\boldsymbol{\hat{\mu}_i}$ is the electric dipole moment of ion $i~(i=1,2)$, $\bf{r}=\bf{r}_2-\bf{r}_1$ is the relative ion position and $\bf{n}=\bf{r}/|r|$. Trapped ions in atomic eigenstates have negligible dipole moments and $\hat{V}_{\mathrm{dd}}$ has no first-order effect. The second-order effect (van der Waals interaction) can be sufficiently strong to cause Rydberg blockade in neutral atom systems with principal quantum number $n\sim50$ separated by a few $\mu \mathrm{m}$ \cite{saffman2010}. However, this interaction is much weaker in Rydberg ion systems; it scales with net core charge as $Z^{-6}$ (for Sr$^+$ with one valence electron $Z=2$) \cite{lebedev1998}. Instead, we achieve a strong first-order interaction by inducing rotating electric dipole moments via a microwave (MW) field.

When two Rydberg states $|s\rangle$ and $|p\rangle$ are coupled by a MW field with Rabi frequency $\Omega_{\mathrm{MW}}$ and detuning $\Delta_{\mathrm{MW}}$ [Fig.~\ref{Fig1} (b)], the eigenstates become $|\pm\rangle = C\left(  \frac{\Delta_{\mathrm{MW}}\pm \sqrt{\Delta_{\mathrm{MW}}^2+\Omega_{\mathrm{MW}}^2}}{\Omega_{\mathrm{MW}}} |s\rangle + |p\rangle \right)$, where $C$ is the normalization constant \cite{li2013}. In our system the electric dipole moments of the dressed states $|\pm\rangle$ rotate with the MW field in the plane perpendicular to the magnetic field, as shown in Fig.~\ref{Fig1}(c) and (d). For two ions, each in state $|r\rangle \equiv |+\rangle$, the dipole-dipole interaction given by Eq.~(\ref{dd_interaction}) yields an energy shift
\begin{equation}
V(\Delta_{\mathrm{MW}},\Omega_{\mathrm{MW}})=\langle rr|\hat{V}_{\mathrm{dd}}|rr\rangle = \frac{1}{4\pi\epsilon_0} \frac{\langle s|\hat{\mu}|p\rangle^2}{r^3} \left( \frac{\Omega_{\mathrm{MW}}^2}{\Delta_{\mathrm{MW}}^2+\Omega_{\mathrm{MW}}^2} \right) \propto \frac{n^4}{Z^2}
\label{interaction_energy}
\end{equation}
with maximum interaction strength $V_{\mathrm{max}} = \frac{1}{4\pi\epsilon_0} \frac{\langle s|\hat{\mu}|p\rangle^2}{r^3}$. For the measurements described in this letter, we use Rydberg states with principal quantum number $n=46$ and ion separation $4.2\,\mu \mathrm{m}$, resulting in $V_\mathrm{max} \simeq 2\pi\times 1.9\,\mathrm{MHz}$. By tuning the ratio between $\Omega_{\mathrm{MW}}$ and $\Delta_{\mathrm{MW}}$ the interaction strength can be varied between $\sim 0$ and $V_\mathrm{max}$. Higher-order terms in $\hat{V}_{\mathrm{dd}}$ can be neglected since the energy splitting between dressed states $\Omega_{\mathrm{MW}}\gg V_{\mathrm{max}}$.

\begin{figure}
	\includegraphics[width=\textwidth]{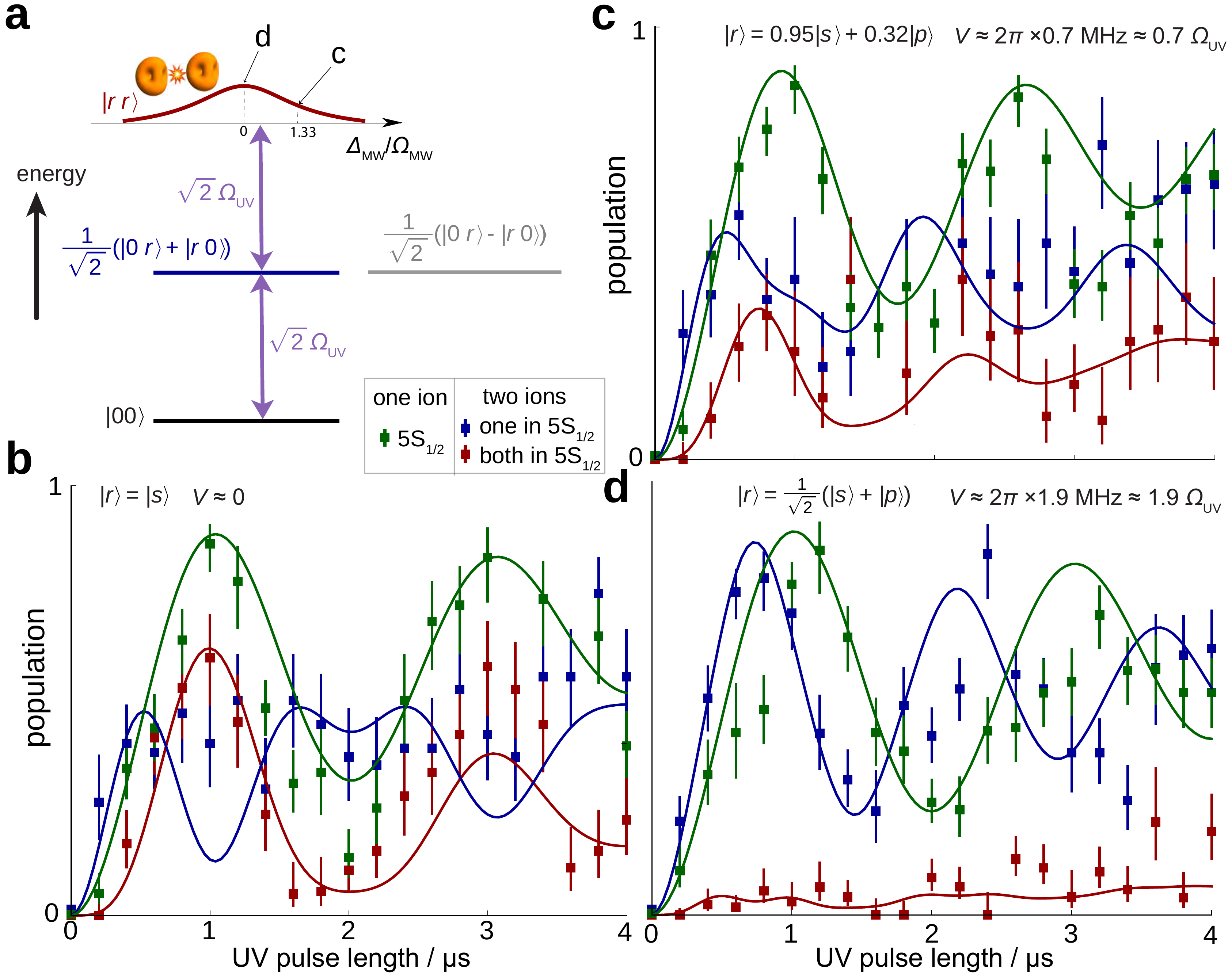}
	\caption{Tunable interaction between Rydberg ions. (a) Two-ion level scheme, where levels are coupled by the two-photon laser field $\Omega_{\mathrm{UV}}=\frac{\Omega_1 \Omega_2}{2\Delta}$, and the energy of state $|rr\rangle$ is shifted by the tunable Rydberg interaction. (b)-(d) Rabi oscillations between $|0\rangle \leftrightarrow |r\rangle$ for one and two ions. From (b)$\rightarrow$(d) the interaction strength $V$ is increased from 0 to $V_\mathrm{max}$ by tuning the MW field. The excitation to $|rr\rangle$ is increasingly suppressed as $|rr\rangle$ is shifted out of resonance from the laser excitation. The y-axes show the populations which is excited to $|r\rangle$ and decays to $5S_{1/2}$. Green data points for one-ion excitation $|0\rangle \leftrightarrow |r\rangle$ are similar for (b)-(d). Red data points show the two-ion excitation probability for $|00\rangle \leftrightarrow |rr\rangle$ which, in (b), is similar to the square of the one-ion (green) data. From (b)$\rightarrow$(d) excitation to $|rr\rangle$ is increasingly suppressed, with Rydberg blockade in (d). The blue data points show the two-ion excitation probability for $|00\rangle \leftrightarrow \frac{1}{\sqrt{2}}(|0r\rangle+|r0\rangle)$; a $\sqrt{2}$-enhanced Rabi frequency compared with the one-ion case (green) is observed in the blockade regime (d). Error bars indicate one standard deviation. Solid lines show results of numerical simulation using no free parameters.}
	\label{Fig2}
\end{figure}

We probe the interaction between Rydberg ions as shown in the Rabi oscillations between $|0\rangle$ and $|r\rangle$ in Fig.~\ref{Fig2}. Either one or two ions are trapped and initialised in $|0\rangle$. A two-photon laser field then couples $|0\rangle \leftrightarrow |r\rangle$. From (b)$\rightarrow$(d) the dipole-dipole interaction strength is increased: this has no effect on the Rabi oscillations when a single ion is trapped (green data), while two-ion oscillations (red data) are suppressed as the interaction shifts the pair state $|rr\rangle$ out of resonance from the laser excitation. When $|rr\rangle$ is far from resonance the population oscillates between two states: $|00\rangle$ and the Bell state $\frac{1}{\sqrt{2}}(|0r\rangle+|r0\rangle)$. This is the blockade regime, which is corroborated by the $\sqrt{2}$ enhancement of two-ion Rabi oscillation frequency (blue data) over the single-ion oscillation frequency (green data) in (d).

\begin{figure}
	\includegraphics[width=0.6\textwidth]{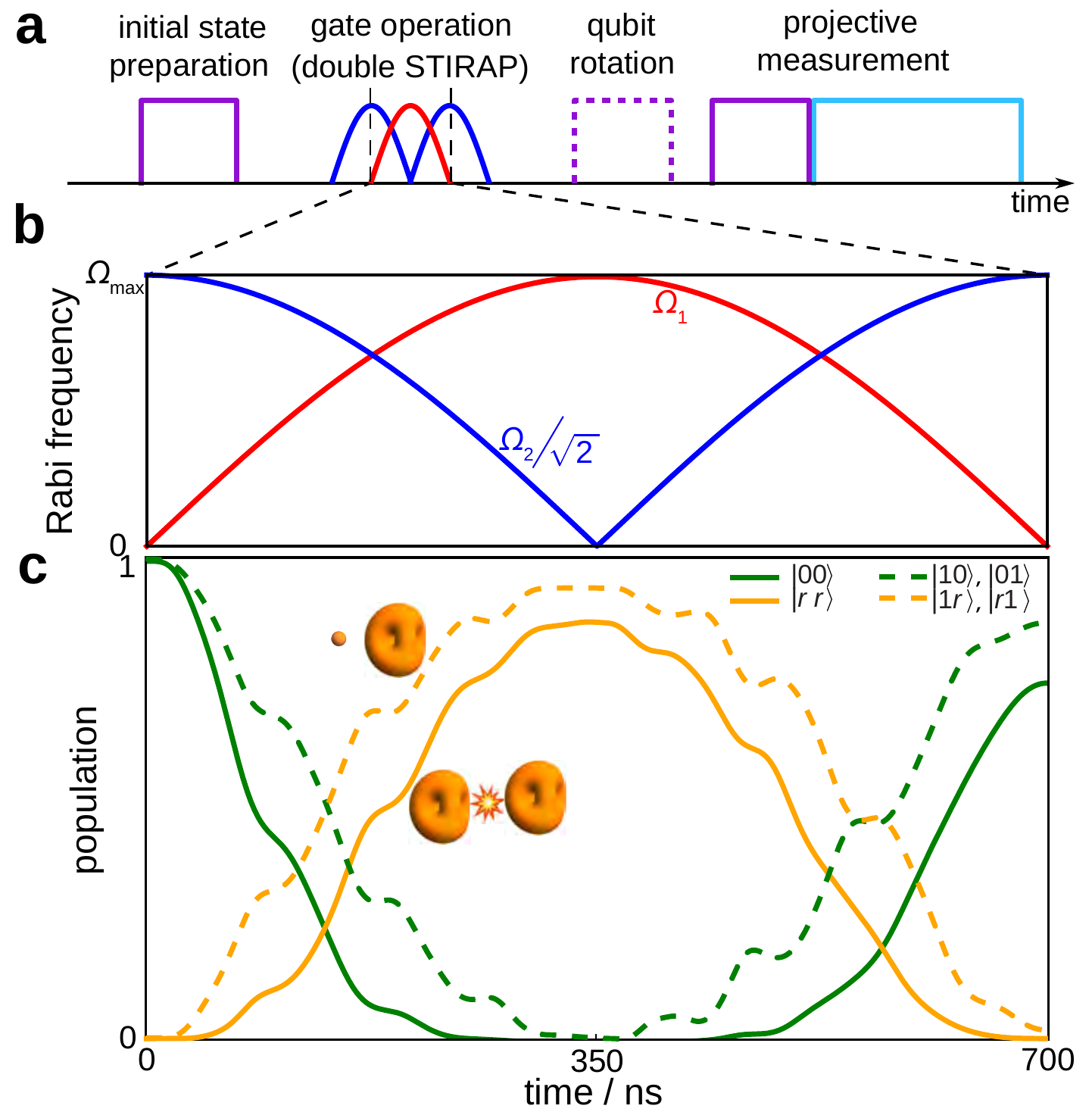}
	\caption{Experimental sequence of the Rydberg-interaction gate. (a) Pulse sequence: two ions are prepared in $\frac{1}{2}\left(|0\rangle+|1\rangle\right)\left(|0\rangle+|1\rangle\right)$, followed by the gate operation involving Rydberg excitation using a double-STIRAP sequence, the qubit rotation is applied for parity-oscillation measurements, and finally the two-ion state is measured using qubit rotation and fluorescence detection. (b) Gate operation: the $|0\rangle \leftrightarrow |e\rangle$ and $|e\rangle \leftrightarrow |r\rangle$ coupling strengths ($\Omega_1$ and $\Omega_2 / \sqrt{2}$) are varied sinusoidally over $700\,\mathrm{ns}$. (c) Simulation of one- and two-ion population dynamics during gate operation, the controlled-phase accumulated by $|00\rangle$ is proportional to area enclosed by the $|rr\rangle$ population curve.}
	\label{Fig3}
\end{figure}

We then use the maximum interaction strength ($V=V_\mathrm{max}$) to implement a $700\,\mathrm{ns}$ controlled phase gate between two ions. The experimental sequence is described in Fig.~\ref{Fig3}. First the two ions are initialized in state $\frac{1}{2}(|00\rangle+|01\rangle+|10\rangle+|11\rangle)$, then the gate operation is applied -- population is transferred from $|0\rangle \rightarrow |r\rangle \rightarrow |0\rangle$, and the Rydberg interaction causes component $|00\rangle$ to acquire phase $\phi$. Finally qubit rotations and projective measurements are used to determine the final two-ion state.

The gate operation consists of a double stimulated Raman adiabatic passage (STIRAP) pulse sequence \cite{rao2014,higgins2017_2}. The three levels $|0\rangle$, $|e\rangle$ and $|r\rangle$ are coupled by two laser fields with coupling strengths $\Omega_1$ and $\Omega_2$. $|0\rangle$ and $|r\rangle$ are resonantly coupled while $|e\rangle$ is detuned by $\Delta$ (see Fig.~\ref{Fig1}). $\Omega_1$ and $\Omega_2$ are gradually changed such that an ion initially in $|0\rangle$ adiabatically follows an eigenstate to $|r\rangle$ and back to $|0\rangle$. An ion initially in $|1\rangle$ is unaffected. For the initial pair states $|11\rangle, |10\rangle$ and $|01\rangle$ the eigenenergies remain zero and no phase is accumulated. From initial state $|00\rangle$ population can be excited to $|rr\rangle$ (provided $\Delta \gtrsim V_{\mathrm{max}}$ \cite{rao2014}), the energy of which is shifted due to the Rydberg interaction. Thus $|00\rangle$ acquires the phase $\phi = V_{\mathrm{max}} \int_{0}^{T} \langle rr|\rho(t)|rr\rangle dt$, with the two-ion density operator $\rho(t)$ and pulse length $T$. We achieve $\phi \simeq \pi$ using sinusoidal profiles for $\Omega_1(t)$, $\Omega_2(t)$ and $T=\frac{8\pi}{3V_\mathrm{max}} \simeq 700\,\mathrm{ns}$. In the ideal case the final target state $\frac{1}{2}(-|00\rangle+|01\rangle+|10\rangle+|11\rangle)$ is maximally entangled.

\begin{figure}
	\includegraphics[width=\textwidth]{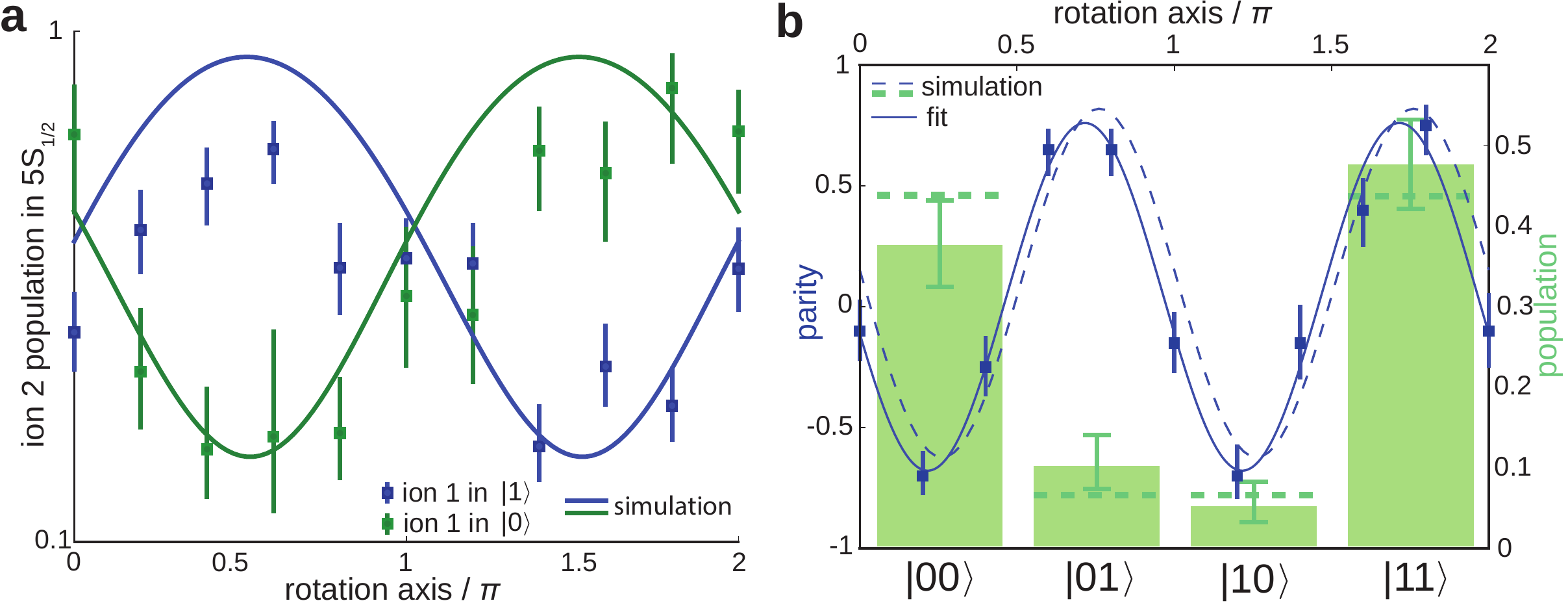}
	\caption{Analysis of the two-ion state after the entangling gate operation. (a) A Ramsey-type experiment measures the relative phase between $|0\rangle_2$ and $|1\rangle_2$ ($\frac{\pi}{2}$-pulse of varied phase applied on ion 2 followed by projection measurement on $\{|0\rangle_2, |1\rangle_2\}$) conditional on the state of ion 1; the $\pi$-phase difference between the cases $\{|0\rangle_1, |1\rangle_1\}$ is consistent with the target entangled state $\frac{1}{2}\left(|0\rangle_1 \left(-|0\rangle_2+|1\rangle_2\right) + |1\rangle_1\left(|0\rangle_2+|1\rangle_2\right)\right)$. (b) $(78\pm 3)\%$ entanglement fidelity is determined by population (green) and parity-oscillation (blue) measurement after rotating the target state to the Bell state $\frac{1}{\sqrt{2}} \left(|00\rangle +i|11\rangle \right)$. Small phase mismatch between simulation and experiment due to minor dynamical phase does not affect entanglement fidelity. Error bars in both (a) and (b) indicate one standard deviation.}
	\label{Fig4}
\end{figure}

The correlation of the final state is measured as follows: the projection of ion 1 on $\{|0\rangle_1, |1\rangle_1\}$ and the phase between $|0\rangle_2$ and $|1\rangle_2$ of ion 2 are measured simultaneously, results are shown in Fig.~\ref{Fig4}(a). The phase between $|0\rangle_2$ and $|1\rangle_2$ of ion 2 is $0$ ($\pi$) when ion 1 is projected onto $|1\rangle_1$ ($|0\rangle_1$) -- this indicates $\phi = \pi$. Entanglement is characterized by parity-oscillation measurements \cite{monz2011,leibfried2005} after rotating the target state to a Bell state [Fig.~\ref{Fig4}(b)]. The coherence and population of the Bell state are measured to be $C=0.72\pm 0.04$ and $P=P_{|00\rangle}+P_{|11\rangle}=0.85\pm 0.04$, which give an entanglement fidelity of $F=\frac{P+C}{2}=0.78\pm 0.03$.

\begin {table}
\caption {Estimation and scaling of gate error from known sources. \footnote{More details about numerical simulation, derivation of the scaling, and estimation of the improved experimental parameters can be found in supplemental information.}}
\label{table} 
\begin{center}
\begin{tabular}{ c|c c c } 
\hline
Error source & This experiment \footnote{Two ions separated by $r=4.2\,\mu \mathrm{m}$, sideband cooled to $\sim 98\%$ population in motional ground state, excited to $n=46$ maximally-dressed Rydberg state with lifetime $\tau_r \simeq 7.8\,\mu \mathrm{s}$ (determined by experiment) and polarisability $\alpha \approx -6\times 10^{-31}\,\mathrm{C^2 m^2 J^{-1}}$, interaction strength $V = V_\mathrm{max}\simeq2\pi \times 1.9\,\mathrm{MHz}$. UV Rabi frequencies $\Omega_1 = 2\pi \times 40\,\mathrm{MHz}$, $\Omega_2 = 2\pi \times 56.5\,\mathrm{MHz}$, and UV laser linewidths $\Gamma_{1,2} \simeq 2\pi \times 10\,\mathrm{kHz}$.} & Scaling & Improved experiment \footnote{We consider a gate acting on two ions in a Doppler-cooled 100-ion crystal, the two ions are separated by $r=2.3\mu \mathrm{m}$, and excited to the zero-polarisability Rydberg state with $n=60$, $\tau_r \simeq 25.5\mu \mathrm{s}$ (natural lifetime), the residual polarisability $|\alpha | < 10^{-34}\,\mathrm{C^2 m^2 J^{-1}}$, $V\simeq2\pi \times 21.9\,\mathrm{MHz}$, $\Gamma_{1,2} \simeq 2\pi \times 1\,\mathrm{kHz}$, $\Omega_1 = 2\pi \times 1000\,\mathrm{MHz}$ and $\Omega_2 = 2\pi \times 1414\,\mathrm{MHz}$.} \\
\hline
Rydberg state decay & $3.5\times 10^{-2}$ & $n^{-7} \cdot r^3$ & $5\times 10^{-4}$ \\ 
Laser linewidth & $\approx 3\times 10^{-2}$ & $\Gamma_l \cdot n^{-4} \cdot r^3$ & $2\times 10^{-4}$ \\ 
Scattering via intermediate state $|e\rangle$ & $8\times 10^{-3}$ & $\Delta^{-2} \cdot n^4 \cdot r^{-3}$ & $8\times 10^{-4}$ \\ 
Transitions within adiabatic basis & $5.5\times 10^{-2}$ & $(\Omega_\mathrm{max}-\Delta /2)^{-2} \cdot n^{4} \cdot r^{-3}$ & $4\times 10^{-4}$ \\
MW power fluctuations & $\approx 10^{-1}$ & $\delta \Omega_\mathrm{MW} \cdot n^{-4} \cdot r^3$ & $< 10^{-5}$\\
Coupling to dressed state $|-\rangle$ & $< 10^{-4}$ & - & $< 10^{-4}$ \\ 
Rydberg state polarisability & $3\times 10^{-3}$ & - & $<10^{-4}$ \\
Coupling to motional modes & $\approx 10^{-4}$ & $N^{1/12}$ & $\approx 10^{-4}$ \\ 
Total error \footnote{Coupling to the quadrupole field of the radio-frequency (rf) trap is not considered; the effect is negligible for $n=46$ state. For higher $n$, it may give a non-negligible energy shift oscillating with the quadrupole field and need to be compensated by modulating the laser and MW frequencies to achieve low error.} & $\approx 23\% $ & - & $ 0.19\% $\\
\hline
\end{tabular}
\end{center}
\end {table}

The contributions of different gate error sources are estimated by numerical simulation (Table~\ref{table}). They can account for the observed gate infidelity. The largest error contributions are technical and can be diminished by improving the MW power stability, increasing the laser intensities and decreasing the laser linewidths. Further, several error contributions depend on the gate time, which can be reduced by using higher Rydberg states which interact more strongly, together with higher laser intensities. In our gate implementation we use sideband cooling to mitigate mechanical effects of Rydberg ion polarisability \cite{higgins2017_1,higgins2019}, which will become unnecessary when we use MW-dressed Rydberg states with zero polarisability \cite{li2013}. In turn, this may allow implementation of the gate in higher-dimensional ion crystals. Importantly, we observe no heating effects of the ion motion after the gate operation in the two-ion crystal. Numerical simulation indicates the gate error induced by mechanical forces between interacting Rydberg ions in a 100-ion crystal is $\sim 10^{-4}$ (using a zero-polarisability MW-dressed state). This simulation and more details about gate errors due to mechanical effects in large ion crystals can be found in the supplemental information.

In summary, we have demonstrated a strong dipole-dipole interaction ($2\pi \times 1.9\,\mathrm{MHz}$) in a trapped ion experiment and observed the Rydberg blockade effect. The interaction can be controlled by a MW field in $\mathrm{ns}$ time scales, which may enable studies of quantum quench dynamics and out-of-equilibrium processes. Moreover, a $700\,\mathrm{ns}$ controlled phase gate has been realised, and the current gate protocol together with technically-achievable experimental parameters may reduce the gate error to 0.2\%. A key feature of the gate is its independence from motional modes. Thus, it should be straightforward to implement this gate in a large ion string, or even in a higher-dimensional ion crystal. This provides a new and promising way to increase both the number of entangling operations within the coherence time and the number of qubits in a trapped ion quantum computer or simulator.

\break
\break

Acknowledgements: We thank Klaus M\o{}lmer for discussions and suggestions for the gate scheme. We thank all members of the ERyQSenS consortium for discussions. This work was supported by the European Research Council under the European Union’s Seventh Framework Programme/ERC Grant Agreement No. 279508, the Swedish Research Council (Trapped Rydberg Ion Quantum Simulator), the QuantERA ERA-NET Cofund in Quantum Technologies (ERyQSenS), and the Knut \& Alice Wallenberg Foundation (“Photonic Quantum Information” and WACQT). I.L. and W.L. acknowledge support from the EPSRC through Grant No. EP/M014266/1 and Grant No. EP/R04340X/1 via the QuantERA project “ERyQSenS”. I.L. also gratefully acknowledges funding through the Royal Society Wolfson Research Merit Award.

Author contributions: G.H., F.P. and M.H. built the experimental system, C.Z. and F.P. set up microwave dressing and improved the UV laser system, A.P. set up ablation loading of ions and the camera software, C.Z. had the idea to combine MW dressing and STIRAP excitation, C.Z. and G.H. carried out the measurements, C.Z. analysed the data, C.Z. and W.L. simulated the results, M.H. designed and administered the experiment, W.L. and I.L. calculated properties of atomic Rydberg states, W.L., I.L. and C.Z. analysed the scaling of the gate error. All authors contributed to discussion and writing the manuscript.

Competing Interests: The authors declare that they have no competing financial interests.

Correspondence: Correspondence and requests for materials should be addressed to C.Z.~(email: chi.zhang@fysik.su.se) or M.H.~(email: markus.hennrich@fysik.su.se).

\newpage

\section*{Supplemental Information: Sub-microsecond entangling gate between trapped ions via Rydberg interaction}

\section{Theory of STIRAP process for two interacting ions}
\label{theory}
For one ion, we have already demonstrated high-efficiency population transfer between $|0\rangle$ and $|r\rangle$ via STIRAP \cite{higgins2017}. As shown in Fig. 1(a) of the main manuscript, when two laser fields $\Omega_1$ and $\Omega_2$ resonantly couple $|0\rangle$ and $|r\rangle$ via the intermediate state $|e\rangle$, the three single-atom eigenstates and eigenenergies (in the rotating-wave frame) are
\begin{eqnarray*}
| \phi_\mathrm{dark} \rangle = \Omega_2 |0\rangle - \Omega_1 |r\rangle, &\quad & E_\mathrm{dark} = 0 \\
| \phi_+ \rangle = \Omega_1 |0\rangle + \left(\Delta + \sqrt{\Omega_1^2 + \Omega_2^2 + \Delta^2}\right)|e\rangle + 	\Omega_2 |r\rangle, &\quad & E_+ = \frac{1}{2} \left(\Delta + \sqrt{\Delta^2 + 4 \Omega_1^2 + 4 \Omega_2^2}\right) \\
| \phi_- \rangle = \Omega_1 |0\rangle + \left(\Delta - \sqrt{\Omega_1^2 + \Omega_2^2 + \Delta^2}\right)|e\rangle + 	\Omega_2 |r\rangle, &\quad & E_- = \frac{1}{2} \left(\Delta - \sqrt{\Delta^2 + 4 \Omega_1^2 + 4 \Omega_2^2}\right)
\end{eqnarray*}
The eigenstate $| \phi_\mathrm{dark} \rangle$ is a superposition of $|0\rangle$ and $|r\rangle$, and the amplitudes and phases of $|0\rangle$ and $|r\rangle$ are determined by the ratio $\Omega_1/\Omega_2$ of the Rabi frequencies. It is called dark state because it does not involve $|e\rangle$. At the beginning of the STIRAP sequence (Fig. 3(b) of the main manuscript), $\Omega_1/\Omega_2=0$ and the initial state $|0\rangle$ is identical to $| \phi_\mathrm{dark} \rangle$. Then the laser fields change gradually and the state follows the dark eigenstate to the final state $| r \rangle$ when $\Omega_2/\Omega_1=0$. To transfer the population back to $|0\rangle$, the reverse pulse sequence is applied. Decoherence is caused by spontaneous decay of the Rydberg state, finite laser linewidth, and further population transfer inefficiency results from transitions between adiabatic eigenstates driven by the time-dependent Hamiltonian. The latter effect is minimised when $\Delta=0$ and the laser Rabi frequencies change sinusoidally.

For two non-interacting ions, the two-ion state follows $ | \phi_\mathrm{dark} \phi_\mathrm{dark} \rangle \equiv |\phi_\mathrm{dark}\rangle_1 |\phi_\mathrm{dark}\rangle_2$ when STIRAP is applied. However, in the presence of the Rydberg interaction, $ | \phi_\mathrm{dark} \phi_\mathrm{dark} \rangle$ is resonantly coupled to $ | \phi_+ \phi_- \rangle$ and $| \phi_- \phi_+ \rangle$, and the coupling strength is
\begin{equation}
\langle \phi_\mathrm{dark} \phi_\mathrm{dark} | \hat{V}_\mathrm{dd}| \phi_+ \phi_- \rangle \approx \frac{V_\mathrm{max}}{2} \left(\frac{\Omega_1 \Omega_2}{\Omega_1^2 + \Omega_2^2}\right)^2
\end{equation}
When both $\Omega_1$ and $\Omega_2$ are non-zero (during STIRAP), the eigenstate has a significant $|e\rangle$-component and decays fast to ground state ($\Gamma_e \simeq 2\pi \times 4.5\,\mathrm{MHz}$). To suppress population $\rho_e$ in $|e\rangle$, a detuning $\Delta \gg V_\mathrm{max}$ needs to be used to separate the eigenenergies of the bright (lossy) eigenstates and the dark eigenstate, since $ \rho_{e,\mathrm{max}} \sim \frac{V_\mathrm{max}^2}{16 \Delta^2}$. The energy shift of the dark state caused by this unwanted coupling is small, $\delta_{E,\mathrm{max}} \sim \frac{V_\mathrm{max}^2}{32 \Delta}$. Additionally, $\Delta \ll \Omega_\mathrm{max}$ is required for adiabaticity.

In the next sections, the STIRAP process with all known error sources is simulated numerically to understand and estimate the gate errors in our experiment.

\section{Numerical simulation}
\label{simulation}
\subsection{Rydberg interaction gate}
During the gate process, six atomic levels are considered, $\{|1\rangle,|0\rangle,|e\rangle,|s\rangle,|p\rangle,|g\rangle\}$, where $|g\rangle$ is the other sublevel of the ground state $5S_{1/2}$. The states $\{|1\rangle,|0\rangle,|e\rangle,|s\rangle,|p\rangle\}$ are described in Fig.~1 in the main manuscript. Spontaneous decay can finish in both sublevels of $5S_{1/2}$ ($|1\rangle$ and $|g\rangle$). The UV laser coupling Hamiltonian is described by 
\begin{equation}
H_L = \frac{1}{2} \Omega_1 (t) (|0\rangle \langle e| + |e\rangle \langle 0|) + \frac{1}{2} \Omega_2 (t) (|e\rangle \langle s| + |s\rangle \langle e|)
\end{equation}
and the Hamiltonian of the MW field is 
\begin{equation}
H_\mathrm{MW} = \frac{1}{2} \Omega_\mathrm{MW} (|s\rangle \langle p| + |p\rangle \langle s|)
\end{equation}
The Hamiltonian describing the interaction between two ions in Rydberg states is 
\begin{equation}
H_I = V_\mathrm{dd} (|sp\rangle \langle ps| + |ps\rangle \langle sp|)
\end{equation}
This yields the total Hamiltonian 
\begin{equation}
H(t) = (H_L(t) + H_\mathrm{MW})_1 \otimes I_2 + I_1 \otimes (H_L(t) + H_\mathrm{MW})_2 + H_I
\end{equation}
The spontaneous decay process from the $|e\rangle$, $|s\rangle$ and $|p\rangle$ to $|1\rangle$ and $|g\rangle$ is described by the collapse operators 
\begin{eqnarray*}
C_{e,1} =\sqrt{\frac{\Gamma_e}{2}} |1\rangle \langle e|,\quad C_{e,g} =\sqrt{\frac{\Gamma_e}{2}} |g\rangle \langle e| \\
C_{s,1} =\sqrt{\frac{\Gamma_s}{2}} |1\rangle \langle s|,\quad C_{s,g} =\sqrt{\frac{\Gamma_s}{2}} |g\rangle \langle s| \\
C_{p,1} =\sqrt{\frac{\Gamma_p}{2}} |1\rangle \langle p|,\quad C_{p,g} =\sqrt{\frac{\Gamma_p}{2}} |g\rangle \langle p|
\end{eqnarray*}

As described in the main text, the initial state is $\frac{1}{2}(|00\rangle+|01\rangle+|10\rangle+|11\rangle)$, and the corresponding Master equation evolution is simulated by the QuTiP \cite{qutip} function Lindblad Master Equation Solver in Python. The interaction strength and the spontaneous decay rates are calculated from atomic matrix elements. The contributions of different error sources are estimated by setting each individual error source to zero and calculate the fidelity difference. More details are discussed in section~\ref{error}.
\subsection{One- or two-ion Rabi oscillations}
The parameters used in the simulation of the Rabi oscillations in Fig.~2 in the main manuscript are as follow: (b) $\Omega_\mathrm{MW}=0$, $\Omega_1= 2\pi\times 18.1\,\mathrm{MHz}$ and $\Omega_2= 2\pi\times 22.0\,\mathrm{MHz}$, no interaction is observed. (c) $\Omega_\mathrm{MW}= 2\pi\times 134\,\mathrm{MHz}$, $\Delta_\mathrm{MW}= 2\pi\times 178\,\mathrm{MHz}$, $\Omega_1= 2\pi\times 20.0\,\mathrm{MHz}$ and $\Omega_2= 2\pi\times 25.6\,\mathrm{MHz}$. (d) $\Delta_\mathrm{MW}=0$, $\Omega_\mathrm{MW}= 2\pi\times 178\,\mathrm{MHz}$, $\Omega_1= 2\pi\times 20.0\,\mathrm{MHz}$ and $\Omega_2= 2\pi\times 28.2\,\mathrm{MHz}$. (b)-(d) UV laser linewidth $\Gamma_l = 50\,\mathrm{kHz}$, intermediate state linewidth $\Gamma_e=2\pi\times 4.5\,\mathrm{MHz}$, Rydberg state lifetimes $\tau_s = 3.5\,\mu \mathrm{s}$ and $\tau_p = 12\,\mu \mathrm{s}$. All parameters are calculated from the corresponding atomic matrix elements (e.g., interaction strength and lifetime at $300\,K$ temperature) or determined by independent measurements (e.g., Rabi frequencies of the laser fields).

\section{Gate error analysis and scaling}
\label{error}
From the numerical simulation described in section~\ref{simulation}, we estimate the contribution from all known error sources. The finite Rydberg state lifetime is an intrinsic error and it contributes $\sim 3.5\%$ error in our experiment. This contribution is proportional to $\Gamma_r T $, with the natural lifetime $\Gamma_r \propto n^{-3}$ and gate time $T \propto V^{-1} \propto n^{-4} r^3$. Therefore $\epsilon_r \sim n^{-7} r^3$ and it can be reduced by using higher Rydberg states and smaller ion-separations which can be achieved by trapping more ions.

The UV laser linewidth of $\Gamma_l \approx 2 \pi \times 10\, \mathrm{kHz}$ contributes $\sim 3\%$ error. This error is $\epsilon_l \sim \Gamma_l T \propto \Gamma_l n^{-4} r^3$. It can be reduced by narrowing the laser linewidths, using higher Rydberg states and smaller ion-separations.

Scattering from the intermediate state $|e\rangle$ is normally not a problem in STIRAP. However, as described in section~\ref{theory}, the dark state has finite $|e\rangle$-population $ \rho_e \propto \frac{V^2}{\Delta^2}$ in presence of strong Rydberg interaction. The contribution to gate error $\epsilon_e \propto \rho_e T \propto \Delta^{-2} n^4 r^{-3}$, in the experiment $\epsilon_e \approx 0.8\%$. It increases with stronger interaction and can only be reduced by large intermediate state detuning $\Delta$.

Transitions between adiabatic basis states result from a too fast variation of the time-dependent Hamiltonian. The speed at which the parameters are varied is $\propto \frac{1}{T}$ and the minimum separation of the adiabatic eigenstates $\approx \left( \Omega - \Delta/2 \right)$. The transition probability, and corresponding error are thus $\epsilon_\Omega \propto \frac{1}{\left( \Omega_\mathrm{max} - \Delta/2 \right)^2 T} \propto \left( \Omega_\mathrm{max} - \Delta/2 \right)^{-2} n^4 r^{-3}$. This gives 5.5\% error in the experiment. To reduce this error, we need high laser Rabi frequencies ($\Omega_\mathrm{max}$) and $\Omega_\mathrm{max} \gg \Delta$. The error may also be reduced by using more complex pulse sequences like composite STIRAP pulses or shortcuts to adiabaticity.

The maximum dressed state $\frac{1}{\sqrt{2}}\left(|s\rangle+|p\rangle\right)$ is used in the experiment. Its polarisability is orders of magnitudes larger than that of the low lying states. Thus the trapping potential of ions in the Rydberg state is different from the ground state \cite{higgins2017}. For each motional Fock state the Rydberg transition frequency is different and cooling to motional ground state is required. The imperfect sideband cooling (98\% in motional ground state) results in an error of 0.3\%. It can be reduced by using a zero-polarisability dressed state with equal and opposite contributions from $|s\rangle$ and $|p\rangle$ (the polarisabilities of $|s\rangle$ and $|p\rangle$ states have opposite signs). The dressed-state polarisability can be measured precisely by spectroscopy and tuned to nearly zero by tuning the MW frequency (the theory value of dressed Rydberg state polarisability is confirmed by our recent experiment \cite{pokorny2019}). For higher Rydberg states ($n \sim 60$) with narrower linewidths, we expect to measure the polarisability to a precision of $10^{-34} \mathrm{C^2 m^2 J^{-1}}$ level and thus achieve a residual polarisability below this value ($10^{-34} \mathrm{C^2 m^2 J^{-1}}$). In that case the error caused by polarisability in a Doppler cooled large ion crystal can be reduced below $10^{-4}$.

In the current experiment, the main error source is the slow MW power fluctuation of $\sim 3\%$. The dressed state energy level depends on the MW Rabi frequency, and thus when averaging over many measurement cycles, this fluctuation contributes $\sim 10\%$ dephasing error. The error is $\epsilon_\mathrm{MW} \propto \delta \Omega_\mathrm{MW} T \propto \delta \Omega_\mathrm{MW} n^{-4} r^3$. As the MW power can be measured to $< 0.01 \%$ precision and controlled instantly, this error can be reduced to $10^{-5}$ level with good power stabilization of the MW field. Alternatively, using a much stronger MW field (in the current experiment we attenuate the MW power by $36\,\mathrm{dB}$) or a bi-chromatic MW field to couple many Rydberg states may result in dressed states that have large dipole moments and are not sensitive to MW power fluctuations.

The error from coupling to motional modes is extremely small and is discussed in the next section (section~\ref{motion}).

Our simulations show that the total gate error can be reduced to less than 0.2\% with technically-achievable parameters. We assume using a high Rydberg state $n = 60$ and improved experimental parameters of $r=2.3\mu \mathrm{m}$, $\Omega_1 = 2\pi\times 1000\,\mathrm{MHz}$, $\Omega_2 = 2\pi\times 1414\,\mathrm{MHz}$ and $\Delta = 2\pi\times 100\,\mathrm{MHz}$. Such small ion-separation can be realised in a 100-ion crystal. In such a large crystal cooling to ground state is challenging and therefore we consider using a zero-polarisability state (the electric dipole moment of this state is $\approx$75\% of the maximumly-dressed state). Higher UV laser Rabi frequencies should be possible: the UV output powers are $60\,\mathrm{mW}$ at $306\,\mathrm{nm}$ and $50\,\mathrm{mW}$ at $243\,\mathrm{nm}$, with beam waists of $\sim 8\,\mu \mathrm{m}$ at the ion position. The highest UV Rabi frequencies measured in our experiment are $\sim 60\,\mathrm{MHz}$, and the focusing on the ions could be improved by more than a factor of 10. Also, the output power of the UV lasers can be improved: a $243\,\mathrm{nm}$ laser with up to $1.4\,\mathrm{W}$ output power in continues-wave mode has been demonstrated \cite{burkley2019}. The $306\,\mathrm{nm}$ laser in our experiment is generated by second-harmonic generation from a $612\,\mathrm{nm}$ laser, which is generated by sum-frequency generation of $1550\,\mathrm{nm}$ and $1010\,\mathrm{nm}$ lasers. The $1.4\,\mathrm{W}$ output power of the $1010\,\mathrm{nm}$ laser is the main limitation, while a system with output power of $165\,\mathrm{W}$ has been demonstrated \cite{upa2014}. Alternatively, when the interaction is sufficiently strong, other gate protocols like a Rydberg blockade gate can be used to achieve high fidelity without using extremely high UV laser powers.

\section{Gate error scaling in large ion crystals}
\label{motion}
Rydberg ions may couple to phonon modes due to non-zero polarisability as well as the strong interaction between Rydberg ions. Both will decrease the gate fidelity. In this section we consider only the latter effect as the former can be mitigated by using MW-dressing to produce a zero-polarisability Rydberg state (as discussed in Section II). We theoretically analyse the scaling of the Rydberg gate errors with the number $N$ of ions. We will show that the Rydberg gate is robust in large ion crystals at intermediate temperatures ($10 \sim 100\mu$K, which can be reached by Doppler Cooling). 


In Rydberg states, the long-range dipole-dipole interaction between the $j$-th and $k$-th ion is
\begin{equation}
\label{eq:expansion}
V_{jk}=\frac{C_3}{|d_j+x_j-d_k-x_k|^3}\approx \frac{C_3}{R_{jk}^3} -\frac{3C_3}{R_{jk}^4}(x_j-x_k),
\end{equation}
where $d_j$ are equilibrium positions of ions and $x_j$ is deviations from the equilibrium position $d_j$. The equilibrium separation between two ions is $R_{jk}=|d_j-d_k|$. We have expanded the interaction potential up to the linear order in the $x_j$. Using the normal coordinate $Q_p$, we have
\begin{eqnarray}
Q_p=\sum_mb^{(p)}_mx_j,\\
x_m=\sum_p b^{(p)}_m Q_p,
\end{eqnarray}
In the interaction picture ($H_0=\sum_p\nu_pa^{\dagger}_pa_p$), the interaction Hamiltonian can be rewritten as
\begin{equation}
\tilde{V}_{jk}\approx W_{jk} -\frac{3W_{jk}}{R_{jk}}\sum_pB_{jk}^{(p)}(a_pe^{-i\nu_p t}+a_p^{\dagger}e^{i\nu_pt}),
\end{equation}
where $W_{jk}=C_3/R_{jk}^3$, $Q_p=l_p(a_p+a_p^{\dagger})$ with $l_p=\sqrt{\hbar/2M\nu_p}$ and $B_{jk}^{(p)}=(b_j^{(p)}-b_k^{(p)})l_p$. $M$ and $\nu_p$ are the atomic mass and mode frequency.
The evolution operator of the system, $U=\mathcal{T}e^{-i\int_0^t V_{jk}(\tau)d\tau}$, can be evaluated explicitly using the Magnus expansion,
\begin{eqnarray}
U&=&\exp\left[-iW_{jk} t-i\left( \frac{3W_{jk}}{R_{jk}}\right)^2\sum_p\left(B_{jk}^{(p)}\right)^2 \frac{\sin \nu_p t -\nu_p t}{\nu_p^2} +\sum_p(f_p a_p^{\dagger} -f_p^* a_p)\right], \nonumber\\
&=& \exp\left[-i\left(W_{jk} -\sum_pg_p^2\nu_p \right) t -i\sum_p g_p^2\sin\nu_pt+\sum_p(f_p a_p^{\dagger} -f_p^* a_p)\right]
\label{eq:evolution}
\end{eqnarray}
with $g_p = \frac{3W_{jk}B_{jk}^{(p)}}{R_{jk}\nu_p}$ and $f_p=g_p(e^{i\nu_pt}-1)$. 

To evaluate how motional modes affect the gate evolution and fidelity, we trace over the phonon part (the last term in the exponent of operator $U$) in the density matrix, and obtain
 the respective coherence factor $C(t)$
\begin{eqnarray}
C(t)=\textrm{Tr}\left[e^{-H_0\beta}e^{\sum_p(f_p a_p^{\dagger} -f_p^* a_p)} \right]
= \exp\left[-G(t)\right],
\end{eqnarray}
where $G(t)=\sum_pg_p^2\coth\frac{\beta\nu_p}{2}(1-\cos\nu_pt)$ is the time-dependent exponent of the coherence factor $C(t)$, and $\beta = 1/k_BT$ with $k_B$ the Boltzmann constant. After tracing, the evolution operator becomes
\begin{equation}
U=C(t)\exp[-i\Phi -i\varphi(t)],
\end{equation}
where $\Phi = \left(W_{jk} -\sum_pg_p^2\nu_p \right) t$ and $\varphi(t) = \sum_p g_p^2\sin\nu_pt$.  

In the ideal situation, a fidelity one gate requires $C(t)=1$, $\Phi = \pi$ and $\varphi = 0$. Deviations from these values cause errors, which are expected to be small, as $g_p\propto l_p/R_{jl}\ll 1$. For example, typically the distance $R_{jk}\sim 4\mu$m and $l_p\sim 10$nm, i.e. $g_p\sim 2\times 10^{-3}$. 

In order to achieve the desired gate one needs to control the interaction $W_{jk}$ (e.g. via distance $R_{jk}$ or Rydberg states), such that $\Phi=\pi$. However this does not guarantee $C(t)=1$, or conversely $G(t)=0$. In the following, we will analyse the exponent $G(t)$ of the coherence factor for an idealised equally spaced crystal and a one-dimension (1D) ion crystal in a harmonic potential.


\subsection{An equally spaced ion crystal}
We start with an idealised crystal, where the ions are equally spaced. The phonon modes of the crystal are obtained by evaluate the Hessian matrix, 
\begin{equation}
\mathcal{H}_{mn}=\left\{\begin{array}{ll}\xi+\sum\limits_{k\neq m}^N\frac{2}{a^3|k-m|^3}, & n=m \\ -\frac{2}{a^3| m-n|^3},&n\neq m \end{array}\right.
\nonumber
\end{equation}
where $a$ is the neighbouring ion spacing.The length and frequency is scaled with respect to $l=\left(\frac{e^2}{4\pi\varepsilon_0 M \omega^2}\right)^{1/3}$ and trap frequency $\omega$. The vibration mode of the ideal crystal can be determined by assuming vibration of each ion is given by $z_m=1/\sqrt{N} e^{i(qma -\nu_qt)}$.  The mode energy is obtained as
\begin{equation}
\nu_q = \sqrt{\xi + \frac{4\textrm{Z}(3)}{a^3} +\sum_m\frac{4\cos (mqa)}{m^3}},
\end{equation}
where $q=2\pi j_q/aN$ ($j_q=0,1,\cdots N-1$) is the wave number. $\textrm{Z}(x)$ is zeta function and $\textrm{Z}(3)\approx 1.2$. 
\begin{figure}
	\includegraphics[width=5.9in]{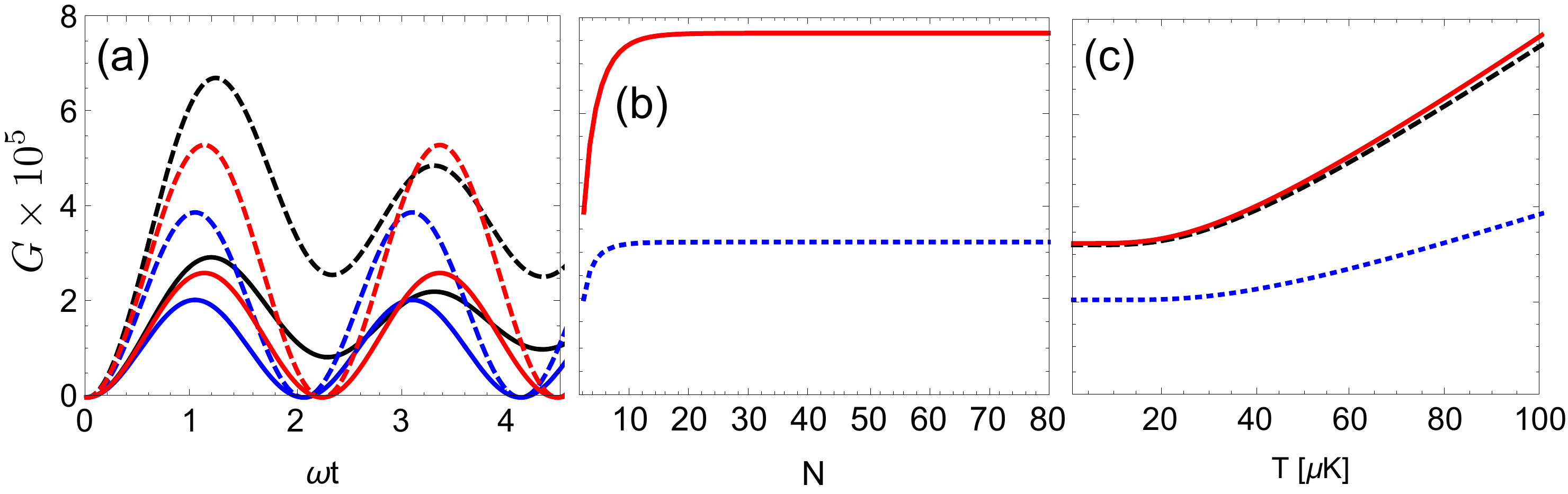}
	\caption{Exponent $G(t)$ for an equally spaced ion crystal. (a) Time evolution of the exponent $G(t)$ at temperature $T=1\,\mu$K (solid) and $T=100\,\mu$K (dashed). We consider 2 (blue), 3 (red), and 20 (black) ions.  (b) $G_{\rm{max}}$ as a function of $N$ at temperature $T=1\,\mu$K (dashed) and $T=100\,\mu$K (solid). All curves saturate when $N>10$. (c) $G_{\rm{max}}$ vs temperature for ion number $N=2$ (dotted), 10 (dashed) and 100 (solid). We also see the saturation of $G_{\rm{max}}$ as a function of ion number $N$. The neighbouring spacing between ions is $5\mu$m. The trapping frequency is $2\pi\times 0.8$ MHz. }
	\label{fig:ring_G}
\end{figure}

Now consider the situation where two neighbouring ions are excited to Rydberg states. The exponent $G(t)$ reads explicitly,
\begin{eqnarray}
G(t)
= \frac{9\hbar}{M}\left(\frac{C_3}{a^4}\right)^2\sum_p\frac{(1-\cos pa)(1-\cos\nu_p t)}{N\nu_p^3}\coth\frac{\beta\nu_p}{2}.
\label{eq:factorG}
\end{eqnarray}
In Fig.~\ref{fig:ring_G}(a), we plot the exponent as a function of time for different temperature $T$ and ion number $N$. The curve shows that the maximal value of $G(t)$ increases slowly as we increase the temperature $T$ and $N$. In the following we analyse the scaling of $G(t)$ with $N$ and $T$ by calculating the upper bound $G_{\rm{max}}=2\sum_p g_p^2\coth\frac{\beta\nu_p}{2}$. This corresponds to all phonon modes being in phase, i.e. $\cos\nu_p=-1$. As shown in Fig.~\ref{fig:ring_G}(b), $G_{\rm{max}}$ saturates when $N>10$, after a rapid growth when $N$ is small. This confirms that the fidelity is not reduced significantly by increasing the number of ions. In Fig.~\ref{fig:ring_G}(c), we show $G_{\rm{max}}$ as a function of temperature. For the idealised crystal, $G_{\rm{max}}$ increases very slowly for intermediate temperatures. $G_{\rm{max}}$ increases quickly when $T>50\,\mu$K. Note that the curves for $N=10$ and 100 are almost identical, i.e. confirming that $G_{\rm{max}}$ saturates with increasing $N$. 

\subsection{1D trapped ion crystal}
\begin{figure}
	\includegraphics[width=6.0 in]{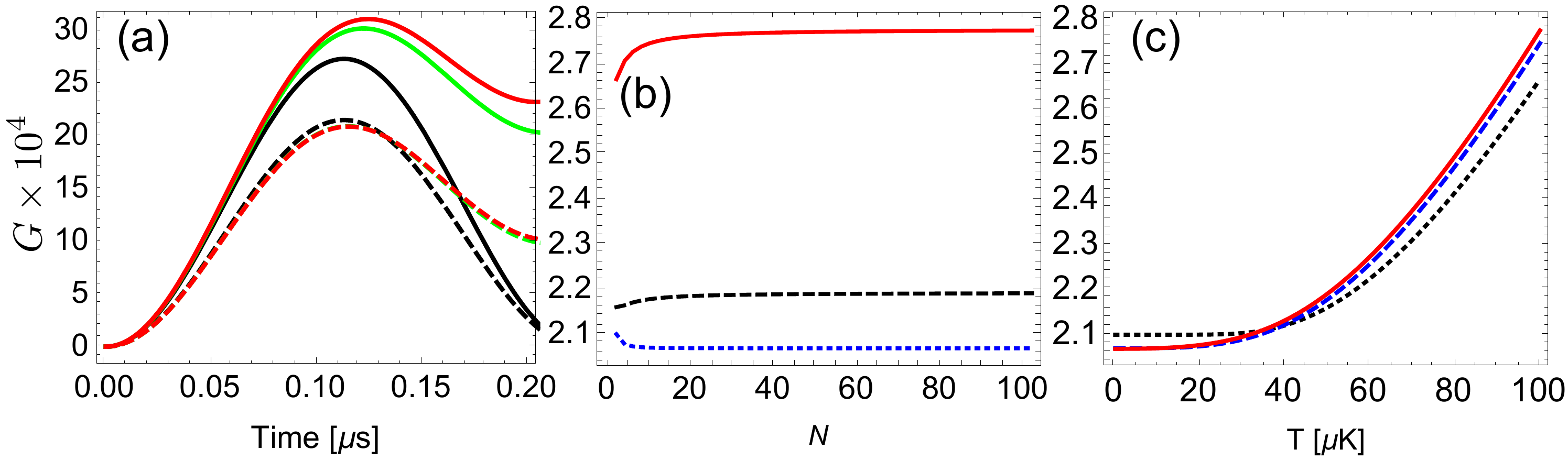}
	\caption{(Color online) Exponent $G(t)$ for a 1D trapped ion crystal. (a) Time evolution of $G(t)$ for 2 (black), 10 (green) and 100 (red) ions. The solid line corresponds to $T=100\,\mu$K and dashed $1\,\mu$K. (b) $G(t_g)$ vs $N$ for temperatures $T$=1 (dotted), 50 (dashed) and 100 $\mu$K  (solid). (c) $G(t_g)$ vs $T$ for $N$ =2 (dotted), 10 (dashed) and 100 (solid). In the calculation, the distance between the middle two ions is 2.3 $\mu$m and the interaction energy is $2\pi \times 21.9$ MHz.}
	\label{fig:trapped_G}
\end{figure}

For 1D ion crystals confined in a harmonic potential, we obtain the coherence factor after numerically solving the corresponding Hessian matrix. As the ion spacing depends on locations of the ions, we implement the entangling gate with two ions in the centre of the crystal. In the analysis, we will fix the spacing between these two ions to be 2.3 $\mu$m even when the length of the ion crystal is varied. For two-ion crystals, this yields the trap frequency $2\pi\times 2.56$ MHz. When we increase the number of ions, the trap frequency that requires to maintain the same ion spacing decreases. 

Time evolution of the exponent $G(t)$ is shown in Fig.~\ref{fig:trapped_G}(a). The oscillation of the exponent depends on ion number and temperature. The increase of $G(t)$ with ion number becomes apparent at higher temperatures. We note that the maximal value of the exponent is larger than the idealised crystal case under same conditions. This is because in the idealised crystal all ions contribute equally [with weight $1/\sqrt{N}$] to the collective phonon modes. In the inhomogeneous crystal the central ions coupled more strongly to the relevant vibrations, i.e. $B_{jk}^{(p)}$ becomes larger in the respective exponent $G(t)$. 

The gate time is largely determined by the interaction between Rydberg ions. With the interaction energy to be $2\pi\times 21.9$ MHz, the resulting gate time is $\tau_g\approx 22.8\,$ns. In Fig.~\ref{fig:trapped_G}(a), the time evolution of the exponent $G(t)$ is shown. Though the value is slightly increased, the qualitative behaviour of the exponent is similar to that of the idealised crystal. 

We shall stress that the gate error shown here is evaluated based on the dipole-dipole interaction between two Rydberg ions. The error is caused by the large gradient of the dipolar interaction [see Eq.~(\ref{eq:expansion})].  However, it is noteworthy that interaction potentials with tailored gradients can be engineered by mixing multiple Rydberg states with MW fields~\cite{Marcuzzi2014}. This would present a further way to reduce mechanical effects and increase the gate fidelity.

\section{Double ionisation by black-body radiation}
Large orbit sizes ($\sim n^2$) in Rydberg states mean that Rydberg electrons are very sensitive to external electromagnetic fields. An immediate effect is that black-body radiation (BBR) will affect the lifetime of Rydberg states, hence the gate fidelity. In general, one would expect that the ionisation rate increases with higher temperature. Here we estimate ionisation rates of the direct BBR photoionisation through a semiclassical approach~\cite{Beterov2009}. The main requirement is that the principal quantum number $n\gg 1$ and angular quantum number $L$ is not very large. This is the situation explored in our experiment, where Rydberg $S$ and $P$ states are excited. Further, taking into account of the fact that the ionisation mainly takes place from the initial state to states close to the double ionisation threshold. One can derive a temperature dependent ionisation rate~\cite{Beterov2009}
\begin{equation}
W_{nL}\approx \frac{\alpha^3k_BT_b}{\hbar\pi^2}\left[\frac{2.8}{n^{7/3}}+\frac{2.09L^2}{n^{11/3}}\right]\log\left[\frac{1}{1-\exp(-\hbar\omega_{nL}/k_BT_b)}\right],
\end{equation}
where $k_B$, $\alpha$ and $\omega_{nL}$ are the Boltzmann constant, fine structure constant and ionisation frequency in SI units, respectively. $T_b$ is temperature of the black-body radiation in Kelvins.

\begin{figure}
	\includegraphics[width=5. in]{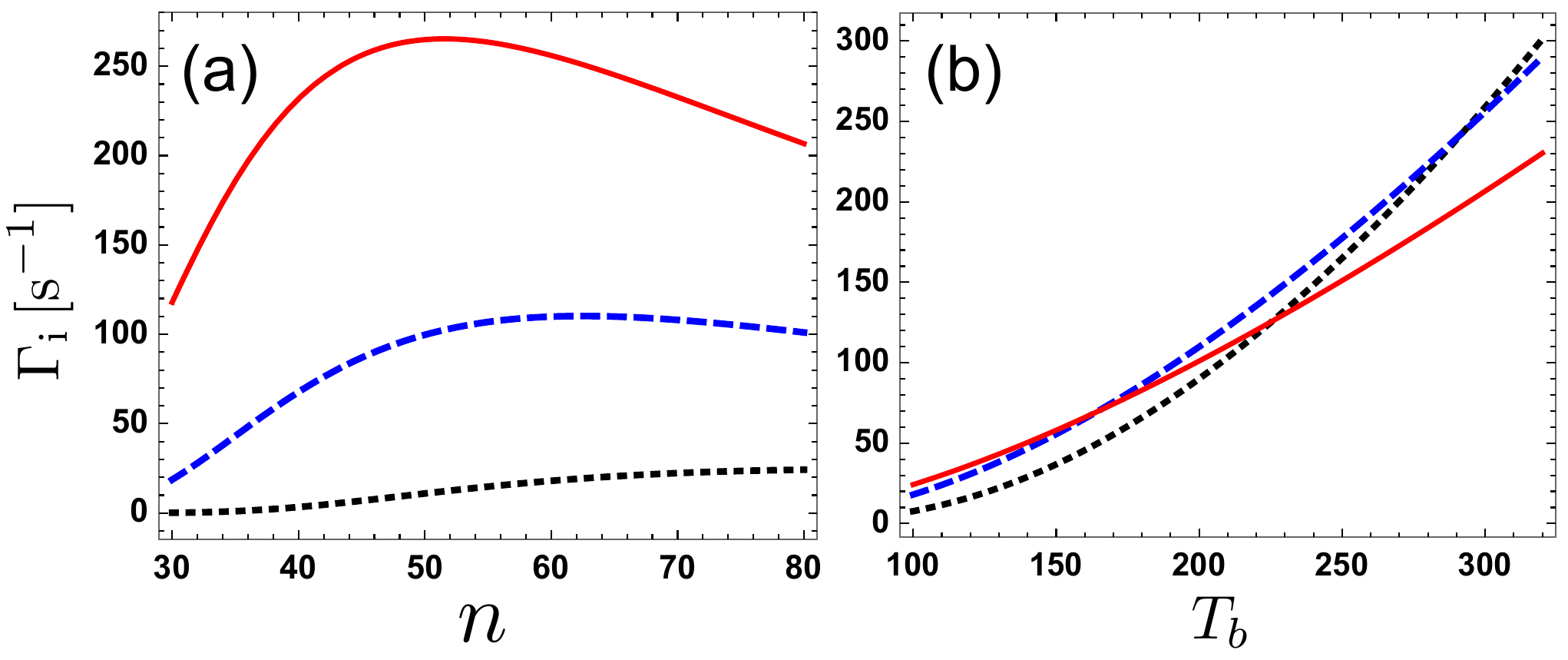}
	\caption{(Color online) Double ionisation rate $\Gamma_i$ of Sr$^+$. (a) $\Gamma_i$ of Rydberg $nS$ state at blackbody temperature $T_b = 100$ K (dotted), $200$ K (dashed) and 300 K (solid). (b) $\Gamma_i$ as a function of the blackbody temperature $T_b$.  At lower temperature, the ionisation rate increases with $n$ gradually. At higher temperature (e.g. $T_b=300$ K), the ionisation rate $\Gamma_i$ depends on $n$ non-monotonically. }
	\label{fig:sr_ionisation}
\end{figure}

As seen in Fig.~\ref{fig:sr_ionisation}(a) the ionisation rate at low temperatures increases with principal quantum number $n$,  (at 300K the BBR double-inonisation rate given by theory can account for ionisation rate we observe in our experiment.). However at higher temperatures the dependence on $n$ is non-monotonic. The ionisation rate first increases and then decreases with $n$. This result is interesting because higher Rydberg states lead to stronger two-body interactions, hence faster gate. In this case, the impact of double ionisation turns out to be weaker [Fig.~\ref{fig:sr_ionisation}(b)]. The data suggest that one can place the ion trap in cryogenic environment to suppress the temperature dependent double ionisation. This allows us to achieve fast Rydberg gates with even higher gate fidelities.


\end{document}